\documentclass[10pt,twocolumn,letterpaper]{article}

\usepackage{cvpr}
\usepackage{times}
\usepackage{epsfig}
\usepackage{graphicx}
\usepackage{amsmath}
\usepackage{amssymb}
\usepackage{caption}
\usepackage{amsmath}
\usepackage{mathrsfs}
\usepackage{multirow}
\usepackage[skip=0pt]{caption}
\usepackage{color}
\usepackage{caption}
\usepackage{subcaption}

\usepackage[pagebackref=true,breaklinks=true,letterpaper=true,colorlinks,bookmarks=false]{hyperref}

 \cvprfinalcopy 

\def\cvprPaperID{2932} 

\ifcvprfinal\fi

\title{Learning Implicit Fields for Generative Shape Modeling}

\author{Zhiqin Chen\\
Simon Fraser University\\
{\tt\small zhiqinc@sfu.ca}
\and
Hao Zhang\\
Simon Fraser University\\
{\tt\small haoz@sfu.ca}
}

\begin{document}


\maketitle
\thispagestyle{empty}


\begin{abstract}
We advocate the use of {\em implicit fields\/} for learning generative models of shapes and introduce an {\em implicit
field decoder\/}, called {\em IM-NET\/}, for shape generation, aimed at improving the visual quality of the generated shapes. An implicit 
field assigns a value to each point in 3D space, so that a shape can be extracted as an iso-surface. IM-NET
is trained to perform this assignment by means of a {\em binary classifier\/}. 
Specifically, it takes a point coordinate, along with a feature 
vector encoding a shape, and outputs a value which indicates whether the point is outside the shape or not.
By replacing conventional decoders by our implicit decoder for representation learning (via IM-AE)
and shape generation (via IM-GAN), we demonstrate superior results for tasks such as generative shape modeling,
interpolation, and single-view 3D reconstruction, particularly in terms of visual quality.
Code and supplementary material are available at \href{https://github.com/czq142857/implicit-decoder}{https://github.com/czq142857/implicit-decoder}.
\end{abstract}

\section{Introduction}
\label{sec:intro}

Unlike images and video, 3D shapes are not confined to one standard representation. Up to date, deep neural networks for 3D shape analysis and synthesis have been developed for voxel grids~\cite{hsp,3DGAN}, multi-view images~\cite{MVCNN}, point clouds~\cite{PCGAN,pointnet}, and integrated surface patches~\cite{atlasnet}. Specific to generative modeling of 3D shapes, despite the many progresses made, the shapes produced by state-of-the-art methods still fall far short in terms of visual quality. This is reflected by a combination of issues including low-resolution outputs, overly smoothed or discontinuous surfaces, as well as a variety of topological noise and irregularities.

\begin{figure}[t!]
   \centering
   \includegraphics[width=1.0\linewidth]{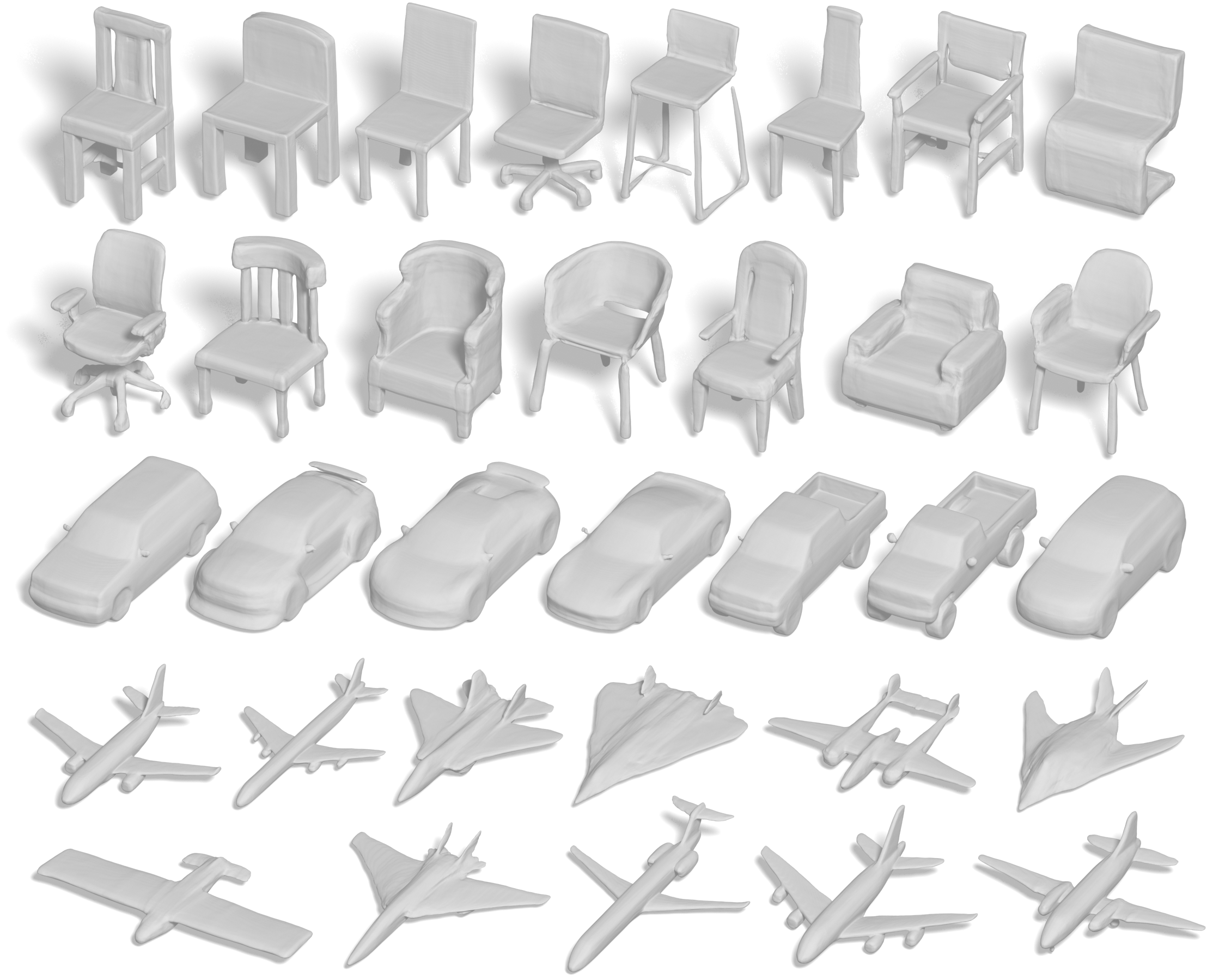}
\vspace{1pt}
   \caption{3D shapes generated by IM-GAN, our {\em implicit field\/} generative adversarial network, which was trained on $64^3$ or $128^3$ voxelized shapes. The output shapes are sampled at $512^3$ resolution and rendered after Marching Cubes.} 
   \label{fig:teaser}
\end{figure}

\begin{figure}[!t]
\begin{center}
   \includegraphics[width=1.0\linewidth]{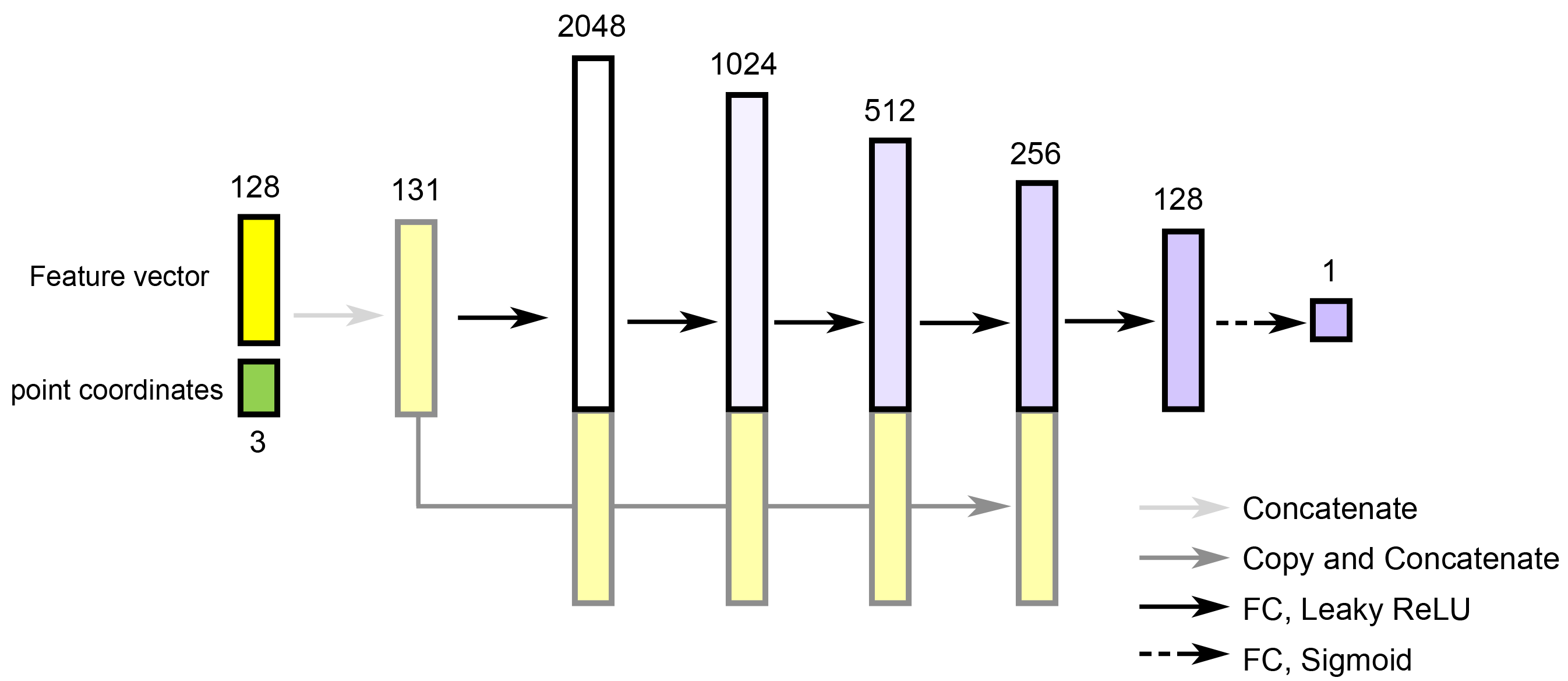}
\end{center}
   \caption{ {\em Network structure of our implicit decoder, IM-NET.} The network takes as input a feature vector extracted by a shape encoder, as well as a 3D or 2D point coordinate, and it returns a value indicating the inside/outside status of the point relative to the shape. The encoder can be a CNN or use PointNET~\cite{pointnet}, depending on the application.}
\label{fig:decoderStructure}
\end{figure}

In this paper, we explore the use of {\em implicit fields\/} for learning deep models of shapes and introduce an {\em implicit field decoder\/} for shape generation, aimed at improving the visual quality of the generated models, as shown in Figure~\ref{fig:teaser}. An implicit field assigns a value to each point $(x, y, z)$. A shape is represented by all points assigned to a specific value and is typically rendered via iso-surface extraction such as Marching Cubes. Our implicit field decoder, or simply implicit decoder, is trained to perform this assignment task, by means of a {\em binary classifier\/}, and it has a very simple architecture; see Figure~\ref{fig:decoderStructure}. Specifically, it takes a point coordinate $(x,y,z)$, along with a feature vector encoding a shape, and outputs a value which indicates whether the point is outside the shape or not. In a typical application setup, our decoder, which is coined {\em IM-NET\/}, would follow an encoder which outputs the shape feature vectors and then return an implicit field to define an output shape.

\begin{figure}[t!]
   \centering
   \includegraphics[width=1.0\linewidth]{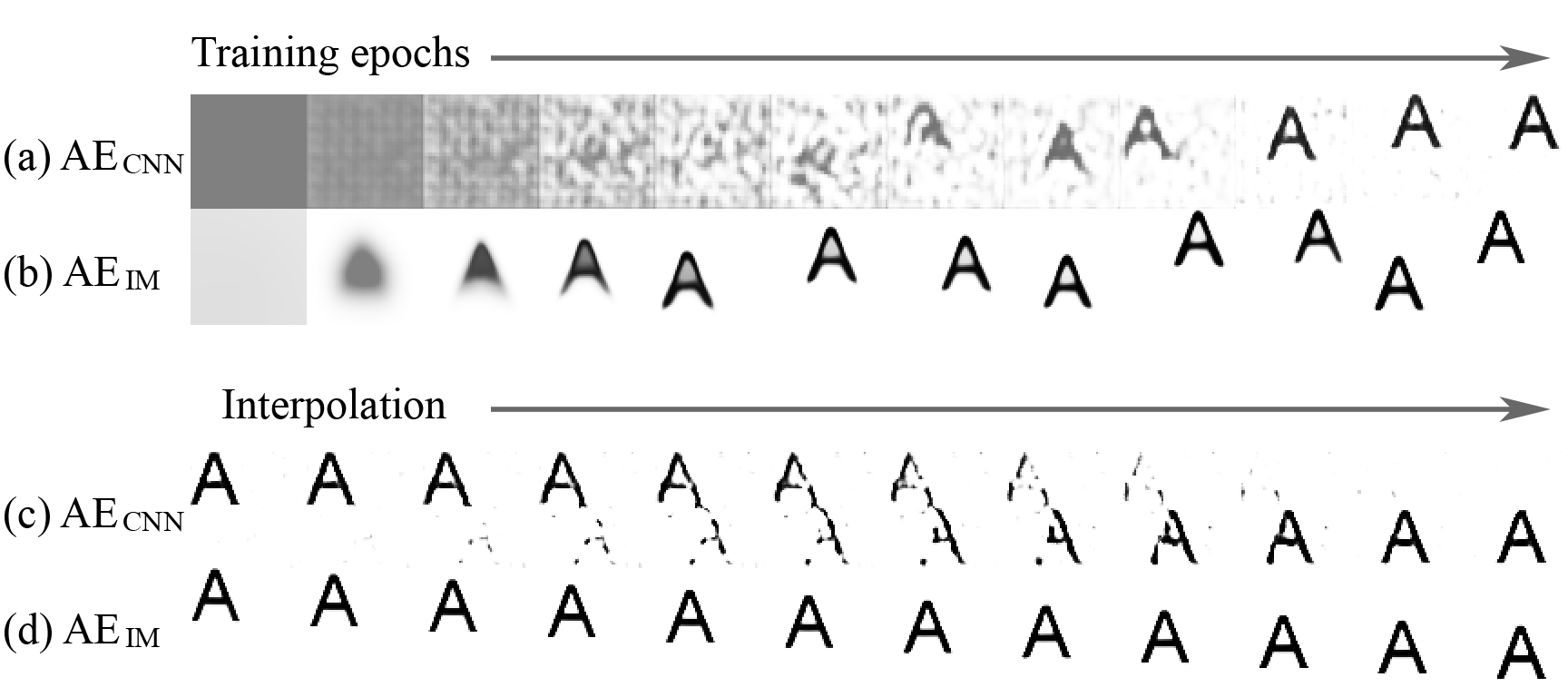}
   \caption{ {\em CNN-based decoder vs.~our implicit decoder.} We trained two autoencoders with CNN decoder ($\text{AE}_{\text{CNN}}$) and our implicit decoder ($\text{AE}_{\text{IM}}$), respectively, on a synthesized dataset of letter A's on white background. The two models have the same CNN encoder. (a) and (b) show the sampled images during AE training. (c) and (d) show interpolation sequences produced by the two trained AEs. See more comparisons in the supplementary material.}
   \label{fig:CNNvsIM}
\end{figure}

Several novel features of IM-NET impact the visual quality of the generated shapes.
First, the decoder output can be sampled at any resolution and is not limited by the
resolution of the training shapes; see Figure~\ref{fig:teaser}. 
%
More importantly, we concatenate point coordinates with shape features, feeding both as 
input to our implicit decoder, which learns the inside/outside status of any point
relative to a shape. In contrast, a classical convolution/deconvolution-based neural network (CNN)
operating on voxelized shapes is typically trained to predict voxels relative to the extent of the
bounding volume of a shape. Such a network learns {\em voxel distributions\/} over the volume, while
IM-NET learns {\em shape boundaries\/}; this is well exemplified in Figure~\ref{fig:CNNvsIM} (top). Experiments show that 
shapes generated by our network possess higher surface quality than results from previous methods, as shown in Figure~\ref{fig:teaser} and results in Section~\ref{sec:results}.

In addition, shape evolution is a direct result of changing the assignments of point coordinates to their inside/outside status and such assignments are precisely what our network, IM-NET, learns. In contrast, convolution kernels compute voxels as weighted averages, where the kernel windows are not ``shape-aware''. Thus a CNN-based decoder typically evolves shape geometries by means of intensity variations; see Figure~\ref{fig:CNNvsIM} (bottom). As a result, our network produces cleaner interpolation results than previous works, even when there are topological changes; see Figure~\ref{fig:GANinterp}.


We embed IM-NET into several contemporary analysis and synthesis frameworks,
including autoencoders (AEs), variational autoencoders (VAEs), and generative adversarial networks (GANs),
by replacing the decoders employed by current approaches with ours, leading to {\em IM-AEs\/} and {\em IM-GANs\/}. This allows assessing the capabilities of our novel decoder for tasks such as shape representation learning, 2D or 3D shape generation, shape interpolation, as well as single-view 3D shape reconstruction. Extensive experiments and comparative studies, both quantitative and qualitative, demonstrate the superiority of our network over previous works, particularly in terms of {\em visual quality\/}.

\section{Related work}
\label{sec:related}

There have been a variety of 3D shape representations for deep learning of shapes, such as voxel grids~\cite{3DR2N2,learningGenerativeEmbeddingCNN,DeepMarchingCubes,3DGAN,SVRGAN}, octrees~\cite{hsp,Octnetfusion,octreeNet,OCNN,AOCNN}, multi-view images~\cite{arsalan2017synthesizing,mutliViewCNN,MVCNN}, point clouds~\cite{PCGAN,pointSetGen,PointCloudTree,pointnet,pointnetplusplus,foldingnet,P2PNET}, geometry images~\cite{DLonGeometryImages,SurfNet}, deformable mesh/patches~\cite{atlasnet,SurfNet,pixel2mesh,foldingnet}, and part-based structural graphs~\cite{GRASS,SCORES}. 
To the best of our knowledge, our work is the first to introduce a deep network for learning implicit fields for generative shape modeling \footnote{Also see \cite{deepSDF,OccNet} for concurrent works that propose similar ideas.}.



With remarkable progress made on generative modeling of images using VAEs~\cite{VAE}, GANs~\cite{WGAN,parzen3,DCGAN}, autoregressive networks~\cite{pixelCNN}, and flow-based models~\cite{glow}, there have been considerably fewer works on generative models of 3D shapes. Girdhar et al.~\cite{learningGenerativeEmbeddingCNN} learned an embedding space of 3D voxel shapes for 3D shape inference from images and shape generation. Wu et al.~\cite{3DGAN} extended GANs from images to voxels and their 3DGAN was trained to generate 3D voxel shapes from latent vectors. Achlioptas et al.~\cite{PCGAN} proposed a latent-GAN workflow that first trains an autoencoder with a compact bottleneck layer to learn a latent representation of point clouds, then trains a plain GAN on the latent code. Common issues with these methods include limited model resolution, uneven and noisy shape surfaces, and inability to produce smooth shape interpolation.

Recently, Li et al.~\cite{GRASS} introduced a part-based autoencoder for 3D shape structures, i.e., a hierarchical organization of part bounding boxes. 
The autoencoder is tuned with an adversarial loss to become generative. Then a separate network is trained to fill in part geometries within the confines of the part bounding boxes. Their method can produce cleaner 3D shapes and interpolation results, mainly owing to the decoupling of structure and geometry generation. However, their network has to be trained by segmented shapes with structural hierarchies. In contrast, our implicit encoder is trained on unstructured voxel shapes.


The output of our decoder IM-NET can be sampled at resolutions higher than that of the training shapes, however, it is not designed for the purpose of
(voxel) super-resolution. There have been works on single image super-resolution using deep networks, \eg,~\cite{dong2014,ledig2017},
which are trained with low- and high-resolution image pairs. Progressive training~\cite{karras2017progressive} is another technique to improve image quality, and we adopt it in our work  
to reduce training times.



Most learning-based methods for single-view 3D reconstruction encode input images with deep convolutional networks, then use an appropriate decoder to reconstruct 3D shapes depending on the shape representations. The most commonly used representations are voxels~\cite{3DR2N2,learningGenerativeEmbeddingCNN,SVRGAN} and point clouds~\cite{pointSetGen,PointCloudTree}. Voxels are natural extensions of image pixels, which allow migrating state-of-the-art techniques from image processing to shape processing. However, voxel representations are usually constrained by GPU memory size, resulting in low-resolution results. Octree representations attempt to fix the memory issues by predicting surfaces in a coarse-to-fine manner~\cite{hsp, AOCNN}.

The recent work by Huang et al.~\cite{DeepVolVideo} also trains a network to perform binary classification, like IM-NET. 
However, the key distinction is that our network assigns inside/outside based on spatial point coordinates. Hence, it learns shape boundaries and effectively, an implicit function that is {\em Lipschitz continuous\/} over point coordinates, i.e., it maps close-by points to similar output values. Moreover, IM-NET can input an arbitrary 3D point and learn a continuous implicit field without discretization. In contrast, their network operates on convolutional features computed over {\em discretized\/} images and learns the inside/outside assignment based on multi-scale image features 
at a point $x$. 

Point clouds can be lightweight on the decoder side by producing few thousand points. However, these points do not provide any surface or topological information and pose a reconstruction challenge. In Deep Marching Cubes, Liao et al.~\cite{DeepMarchingCubes} proposed a differentiable formulation of marching cubes 
to train an end-to-end 3D CNN model for mesh reconstruction from point clouds. However, the resulting mesh still shares common issues with other CNN-based networks, \eg, low resolution ($32^3$) and topological noise.

Some other works~\cite{atlasnet, SurfNet} deform a surface template (\eg, square patches or a sphere) onto a target shape. But many shapes cannot be well-represented by a single patch, 
while outputs from multi-patch integrations often contain visual artifacts due to gaps, foldovers, and overlaps. 
%
In Pixel2Mesh, Wang et al.~\cite{pixel2mesh} used a graph-based CNN~\cite{graphNN} to progressively deform an ellipsoid template to fit an image. The end-to-end network directly generates  meshes but the results tend to be overly smoothed, capturing only low-frequency features while restricted to sphere topology.
%

\section{Implicit decoder and shape generation}
\label{sec:method}


An implicit field is defined by a continuous function over 2D/3D space. Then a mesh surface can be reconstructed by finding the zero-isosurface of the field with methods such as Marching Cubes~\cite{marchingcubes}. In our work, we consider using neural networks to describe a shape in such an implicit way. For a closed shape, we define the inside/outside field $\mathcal{F}$ of the shape by taking the sign of its signed distance field:
\begin{align}
\mathcal{F}(p) = \left\{ \begin{array}{ll}
0 & \textrm{if point $p$ is outside the shape,}\\
1 & \textrm{otherwise.}
\end{array} \right.
\end{align}
Assume the in-out field is restricted in a unit 3D space, we attempt to find a parameterization $f_\theta(p)$ with parameters $\theta$ that maps a point $p \in [0,1]^3$ to $\mathcal{F}(p)$. This is essentially a binary classification problem, which has been studied very well. Multi-layer perceptrons (MLPs) with rectified linear unit (ReLU) nonlinearities are ideal candidates for such tasks. With sufficient hidden units, MLP family is able to approximate the field $\mathcal{F}$ within any precision. This is a direct consequence of the universal approximation theorem~\cite{universalapprox}. Notice that using MLPs also gives us a representation which is continuous across space, so that the mesh can be recovered by taking the $k$-isosurface of the approximated field, where $k$ is an appropriate threshold.

\subsection{Data preparation}

The training of such implicit model needs point-value pairs. It is natural to first voxelize or rasterize the shape for the sake of convenience and uniform sampling. For 3D shapes, we use the same technique as in Hierarchical Surface Prediction (HSP)~\cite{hsp} to get the voxel models in different resolutions ($16^3$, $32^3$, $64^3$, $128^3$). We sample points on each resolution in order to train the model progressively.

A naive sampling would take the center of each voxel and produce $n^3$ points. A more efficient approach when dealing with shapes is to sample more points near shape surfaces and neglect most points far away, leading to roughly $O(n^2)$ points. To compensate for the density change, we assign a weight $w_p$ to each sampled point $p$, representing the inverse of the sampling density near $p$. The implementation of such a sampling method is flexible and varies according to resolution and shape category; more details can be found in the supplementary material.
Most 2D shapes are already rasterized into images. For simplicity, we apply the naive sampling approach for images with an appropriate threshold to determine whether a pixel belongs to the shape.

\subsection{Network structure of IM-NET}

Our model is illustrated in Figure~\ref{fig:decoderStructure}. The skip connections (copy and concatenate) in the model can make the learning progress faster in the experiments. They can be removed when the feature vector is long, so as to prevent the model from becoming too large. The loss function is a weighted mean squared error between ground truth labels and predicted labels for each point. Let $S$ be a set of points sampled from the target shape, we have:
\begin{equation}
\mathcal{L}(\theta) = \frac{ \sum_{p \in S}{ |f_\theta(p) - \mathcal{F}(p)|^2 \cdot w_p } }{ \sum_{p \in S}{w_p} }
\end{equation}

\subsection{Shape generation and other applications}

Our implicit field decoder, IM-NET, can be embedded into different shape analysis and synthesis frameworks to support various applications. In this paper, we demonstrate shape autoencoding, 2D and 3D shape generation, and single-view 3D reconstruction. Due to page limit, the models are briefly introduced here. Detailed structures and hyperparameters can be found in the supplementary material.

For auto-encoding 3D shapes, we used a 3D CNN as encoder to extract $128$-dimensional features from $64^3$ voxel models. We adopt progressive training techniques, to first train our model on $16^3$ resolution data, then increase the resolution gradually. Notice that the structure of the model does not change when switching between training data at different resolutions, thus higher-resolution models can be trained with pre-trained weights on low-resolution data. In the experiments, progressive training can stabilize training process and significantly reduce training time.

For 3D shape generation, we employed latent-GANs~\cite{PCGAN, trainingGAN} on feature vectors learned by a 3D autoencoder. We did not apply traditional GANs trained on voxel grids since the training set is considerably smaller compared to the size of the output. Therefore, the pre-trained AE would serve as a means for dimensionality reduction, and the latent-GAN was trained on high-level features of the original shapes. We used two hidden fully-connected layers for both the generator and the discriminator, and the Wasserstein GAN loss with gradient penalty~\cite{WGAN, WGANGP}. In generative models for 2D shapes, we used the same structure as in the 3D case, except the encoder was 2D CNN and the decoder took a 2D point as input. We did not apply progressive training for 2D shapes since it is unnecessary when the images are small.

For single-view 3D reconstruction (SVR), we used the ResNET~\cite{resnet} encoder to obtain $128$-D features from $128^2$ images. We followed the idea from AtlasNET~\cite{atlasnet} to first train an autoencoder, then fix the parameters of the implicit decoder when training SVR. In our experiments, we adopted a more radical approach by only training the ResNET encoder to minimize the mean squared loss between the predicted feature vectors and the ground truth. This performed better than training the image-to-shape translator directly, since one shape can have many different views, leading to ambiguity. Pre-trained decoders provide strong priors that can not only reduce such ambiguity, but also shorten training time, since the decoder was trained on unambiguous data in the autoencoder phase and encoder training was independently from the decoder in SVR phase.

We reconstructed the 3D meshes by Marching Cubes, and 2D images by sampling a grid of points then optionally applying thresholding to obtain binarized shapes.

\section{Results and evaluation}
\label{sec:results}

In this section, we show qualitative and quantitative results on various tasks using our implicit decoder, IM-NET, and compare them with state-of-the-art approaches. We used the dataset provided by~\cite{hsp}, which contains $256^3$-voxelized and flood-filled 3D models from ShapeNet Core dataset (v1)~\cite{shapenet}, and the corresponding rendered views. To compare with other methods that output point clouds, we first used Marching Cubes to obtain meshes from the $256^3$-voxelized models, then used Poisson-disk Sampling~\cite{PoissonDiskSampling} to obtain 10000 points. This gave us point clouds with only points on the surfaces of the meshes. We evaluated our method, and others, on five representative categories: plane, car, chair, rifle, and table. 
These categories contain 4,045, 7,497, 6,778, 2,373, and 8,509 3D shapes, respectively.

\subsection{Quality metrics}

In our experiments, both qualitative (via visual examination) and quantitative evaluations are provided.
Specific to shapes, most evaluation metrics for encoding and reconstruction are based on point-wise
distances, \eg, chamfer distance (CD), or global alignment,
\eg, mean squared error (MSE) and intersection over union (IoU) on voxels. However, these may not be the best
{\em visual\/} similarity or quality metrics. For example, slightly adjusting the angles between the legs of
a chair and its seat 
may be barely perceptible, compared to removing one shape part, yet the latter could lead to a lower CD or IoU. 
Past works, \eg,~\cite{sorkine2014}, have shown that low-frequency displacements (\eg, bending a leg) in
shapes are less noticeable than high-frequency errors over local surface characteristics such as
normals and curvatures. Metrics such as MSE, CD, and IoU do not account for visual quality of
the object surfaces. 

A less frequently used visual similarity metric in the computer vision community, the light field descriptor (LFD)~\cite{LFD}, 
has been widely adopted in computer graphics. Inspired by human vision system, LFD
considers a set of rendered views of a 3D shape from various camera angles. Each projected image is then
encoded using Zernike moments and Fourier descriptors for similarity comparisons.

\subsection{Auto-encoding 3D shapes}

\begin{table}[!t]
\begin{center}
\begin{tabular}{l|c|c|c|c|c}
\hline
  & Plane & Car & Chair & Rifle & Table \\
\hline\hline
CNN64-MSE & {\bf 1.47} & {\bf 4.37} & {\bf 7.76} & {\bf 1.62} & {\bf 5.80} \\
IM64-MSE & 2.14 & 4.99 & 11.43 & 1.91 & 10.67 \\
\hline
CNN64-IoU & {\bf 86.07} & {\bf 90.73} & {\bf 74.22} & {\bf 78.37} & {\bf 84.67} \\
IM64-IoU & 78.77 & 89.26 & 65.65 & 72.88 & 71.44 \\
\hline
CNN64-CD & {\bf 3.51} & 5.31 & {\bf 7.34} & {\bf 3.48} & {\bf 7.45} \\
IM64-CD & 4.22 & {\bf 5.28} & 8.96 & 3.78 & 12.05 \\
IM256-CD & 4.23 & 5.44 & 9.05 & 3.77 & 11.54 \\
\hline
CNN64-LFD & 3,375 & 1,323 & 2,555 & {\bf 3,515} & {\bf 1,824} \\
IM64-LFD & 3,371 & 1,190 & 2,515 & 3,714 & 2,370 \\
IM256-LFD & {\bf 3,236} & {\bf 1,147} & {\bf 2,453} & 3,602 & 2,201 \\
\hline
\end{tabular}
\end{center}
\caption{ {\em 3D reconstruction errors.} CNN and IM represent CNN-AE and IM-AE, respectively, with $64$ and $256$ indicating sampling resolutions. The mean is taken over the first 100 shapes in each tested category. MSE is multiplied by $10^3$, IoU by $10^2$, and CD by $10^4$. LFD is rounded to integers. Better-performing numbers are shown in boldface.}
\label{table:AEresults}
\end{table}

\begin{figure}[!t]
\begin{center}
   \includegraphics[width=1.0\linewidth]{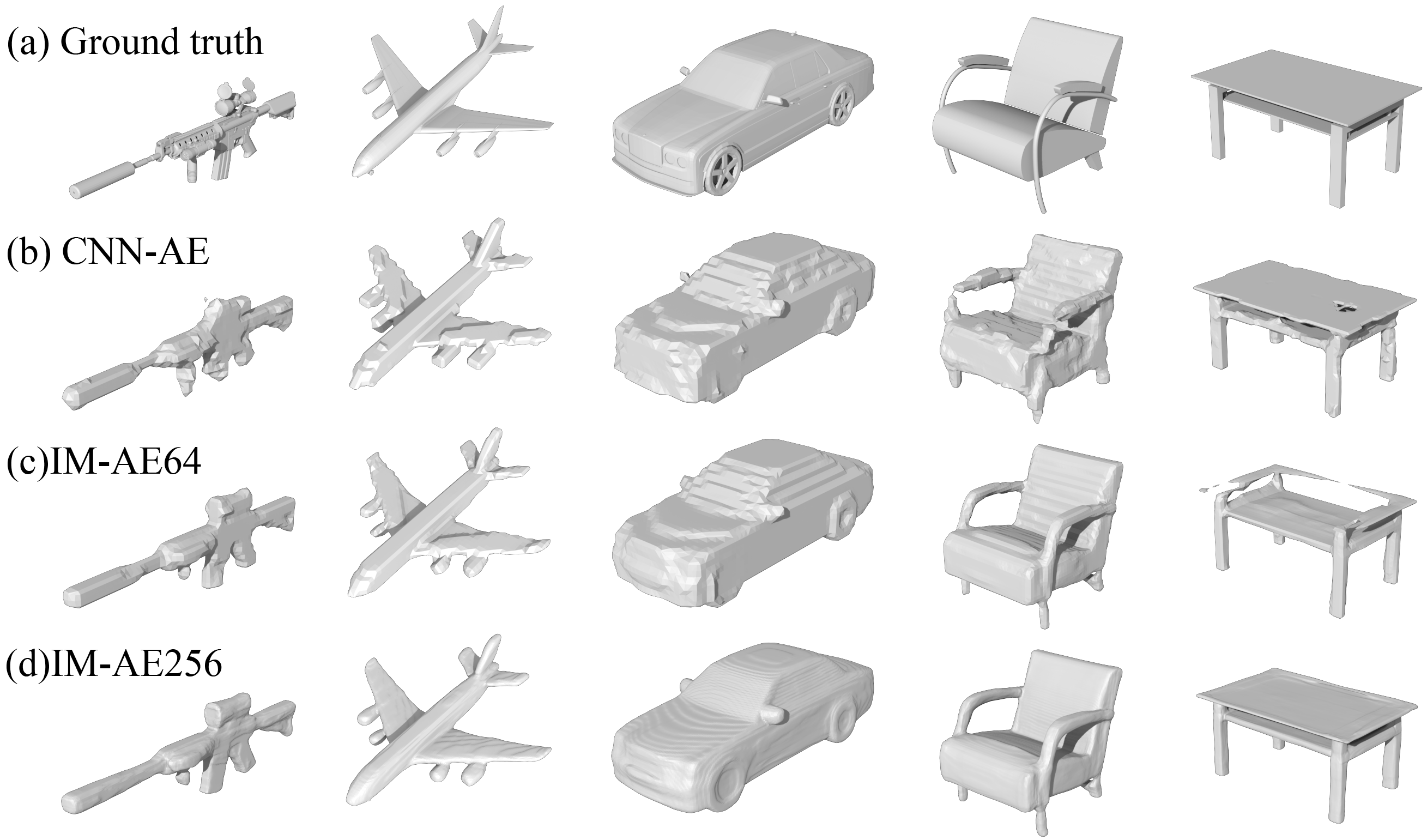}
\end{center}
   \caption{ {\em Visual results for 3D reconstruction.} Each column presents one example from one category. IM-AE64 is sampled on $64^3$ resolution and IM-AE256 on $256^3$. All results are rendered using the same Marching Cubes setup.}
\label{fig:AEvisual}
\end{figure}

We first compare IM-NET with CNN decoders. For each category, we sorted the shapes by name and used the first 80\% as training set and the rest for testing. We trained one model with our implicit decoder (IM-AE) and another with 3D CNN decoder (CNN-AE) on each category. The 3D CNN decoder is symmetric to the 3D CNN encoder (details can be found in the supplementary material). Both models had the same encoder structure and were trained on $64^3$ resolution for the same number of epochs between 200 and 400, depending on the size of the dataset.

Table~\ref{table:AEresults} evaluates reconstruction results using several common evaluation metrics: MSE, IoU,  symmetric Chamfer distance (CD), and LFD. MSE and IoU were computed against the ground truth $64^3$ voxel models. For CD and LFD, we obtained meshes from the output voxel models by Marching Cubes with threshold 0.5. We sampled 2,048 points from the vertices for each output mesh and compare against ground truth point clouds to compute CD. Note that CNN-AE has fixed output size ($64^3$), yet our implicit model can be up-sampled to arbitrarily high resolution by adjusting the sampling grid size. In Table~\ref{table:AEresults}, IM256 is the same IM-AE model but sampled at $256^3$.

Although CNN-AE beats IM-AE in nearly all five categories in terms of MSE, IOU, and CD, visual examination clearly reveals that IM-AE produces better results, as shown in Figure~\ref{fig:AEvisual}; more such results are available in the supplementary material. This validates that LFD is a better visual similarity metric for 3D shapes. On one hand, the movement of some parts, for example, table boards, may cause significant MSE, IOU and CD changes, but bring little visual changes; on the other hand, legs are usually thin, so that missing one leg may cause minor MSE, IOU and CD changes, but can bring significant visual changes. As remarked above, MSE, IOU and CD do not capture well surface quality: a smooth but not exactly aligned surface might have poorer evaluation results than an aligned jagged surface. The situation is better when using LFD. However, since LFD only renders the silhouette of the shape without lighting, it can only capture surface condition on the edge of the silhouette. We expect better evaluation metrics to be proposed in the future, and for the following experiments, we use LFD as our primary evaluation metric.

Note that in the table example in Figure~\ref{fig:AEvisual}, a $64^3$ resolution represents an under-sampling, while going up to $256^3$ reveals more details. This indicates that our generative model is able to generate a table board thinner than the resolution of the training data, which shows that the model learned the implicit field from the whole shape in the space rather than merely learning voxel distributions.

\subsection{3D shape generation and interpolation}

\begin{figure}[!t]
\begin{center}
   \includegraphics[width=1.0\linewidth]{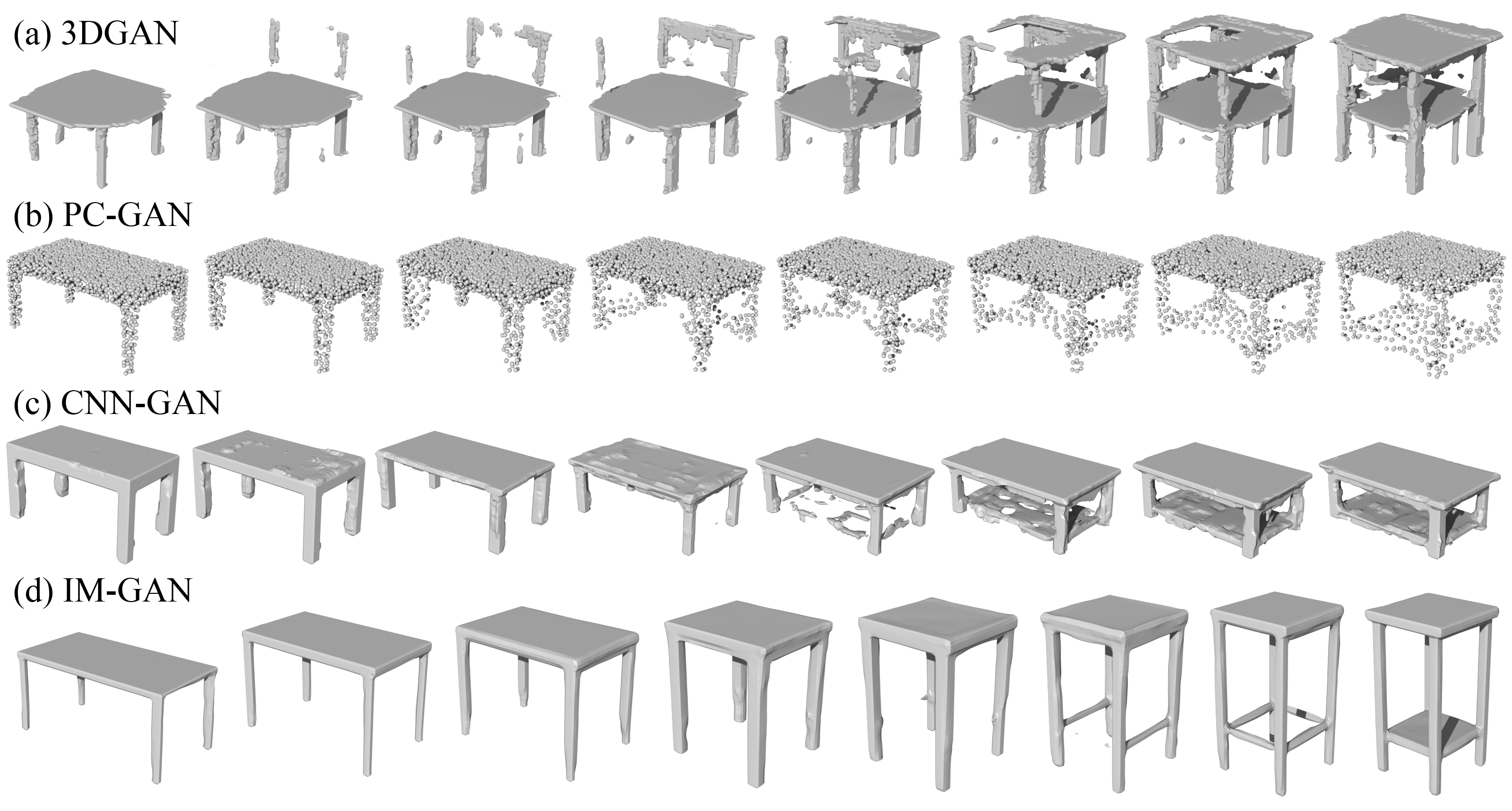}
\end{center}
   \caption{ {\em 3D shape interpolation results.} 3DGAN, CNN-GAN, and IM-GAN are sampled at $64^3$ resolution to show the smoothness of the surface is not just a matter of sampling resolution. Notice that the morphing sequence of IM-GAN not only consists of smooth part movements (legs, board), but also handles topology changes.}
\label{fig:GANinterp}
\end{figure}

\begin{table*}[!t]
\begin{center}
\begin{tabular}{l|c|c|c|c|c|c|c|c}
\hline
 & & Plane & Car & Chair & Rifle & Table & Average w/o planes & Average \\
\hline\hline
\multirow{2}{*}{COV-LFD (\%)} & 3DGAN~\cite{3DGAN} &  & 12.13 & 25.07 & 62.32 & 18.80 & 29.58 &  \\
& PC-GAN~\cite{PCGAN} & {\bf 73.55} & 61.40 & 70.06 & 61.47 & 77.50 & 67.61 & 68.80  \\
& CNN-GAN & 69.22 & {\bf 73.00} & {\bf 77.73} & 61.26 & 83.73 & 73.93 & 72.99  \\
& IM-GAN & 70.33 & 69.33 & 75.44 & {\bf 65.26} & {\bf 86.43} & {\bf 74.12} & {\bf 73.36}  \\
\hline
\multirow{2}{*}{MMD-LFD} & 3DGAN~\cite{3DGAN} &  & 1,993 & 4,365 & 4,476 & 5,208 & 4,010 &  \\
& PC-GAN~\cite{PCGAN} & 3,737 & 1,360 & 3,143 & 3,891 & 2,822 & 2,804 & 2,991  \\
& CNN-GAN & 3,745 & 1,288 & 3,012 & 3,819 & 2,594 & 2,678 & 2,892  \\
& IM-GAN & {\bf 3,689} & {\bf 1,287} & {\bf 2,893} & {\bf 3,760} & {\bf 2,527} & {\bf 2,617} & {\bf 2,831}  \\
\hline
\end{tabular}
\end{center}
\vspace{-5pt}
\caption{ {\em Quantitative evaluation of 3D shape generation.} LFD is rounded to integers. See texts for explanation of the metrics.} 
\label{table:GANresults}
\end{table*}

\begin{figure*}[!t]
\begin{center}
   \includegraphics[width=\linewidth]{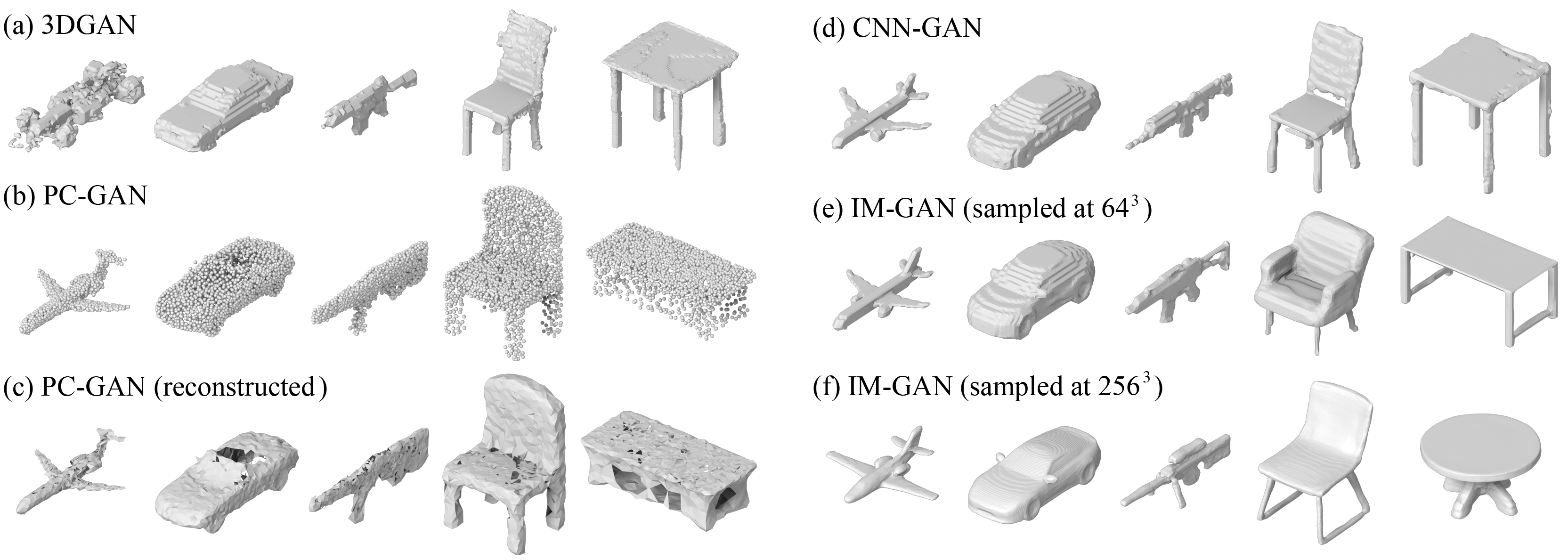}
\end{center}
\vspace{-5pt}
   \caption{ {\em 3D shape generation results.} One generated shape from each category is shown for each model; more results are available in the supplementary material. Since the trained models of 3DGAN did not include plane category, we fill the blank with another car. The ball-pivoting method~\cite{ballPivoting} was used for mesh reconstruction (c) from PC-GAN results (b).}
\label{fig:GANvisual}
\end{figure*}

Next, we assess and evaluate improvements made by our implicit decoder for generative modeling of 3D shapes. We trained latent-GANs on both CNN-AE and IM-AE to obtain CNN-GAN and IM-GAN. We also compared our results with two state-of-the-art methods, 3DGAN~\cite{3DGAN} and the generative model for point clouds in~\cite{PCGAN} (PC-GAN). For 3DGAN, we used the trained models made available online by the authors. PC-GAN was trained using latent WGAN~\cite{PCGAN}. The autoencoder of PC-GAN was trained on our aforementioned point cloud data for 400 epochs for each category. PC-GAN, CNN-GAN, and IM-GAN were trained on the training split of the dataset for 10,000 epochs. 3DGAN was not trained using train/test split~\cite{3DGAN}.

To compare the generative schemes, we adopted the evaluation metrics from~\cite{PCGAN}, but replaced CD or EMD by LFD. Suppose that we have a testing set $G$ and a sample set $A$, for each shape in $A$, we find its closest neighbor in $G$ using LFD, say $g$, and mark $g$ as ``matched''. In the end, we calculate the percentage of $G$ marked as ``matched'' to obtain the coverage score (COV-LFD) that roughly represents the diversity of the generated shapes. However, a random set may have a high coverage, since matched shapes need not be close. Therefore, we match every shape in $G$ to the one in $A$ with the minimum distance and compute the mean distances in the matching as Minimum Matching Distance (MMD-LFD). Ideally, a good generative model would have higher COV-LFD and lower MMD-LFD values.

We first sampled shapes using the subject generative model to obtain $A$, where the number of sampled shapes is five times the number of shapes in the testing split ($G$) of that category. For PC-GAN, we employed the ball-pivoting method~\cite{ballPivoting} to reconstruct the shape surface, while for all the other generative models, we used Marching Cubes. IM-GAN was sampled at $64^3$ in quantitative evaluation.

Quantitative and qualitative evaluations are shown in Table~\ref{table:GANresults} and Figure~\ref{fig:GANvisual}, respectively. Overall, IM-GAN performs better on both COV-LFD and MMD-LFD. More importantly, IM-GAN generates shapes with better visual quality compared to other methods, in particular, with smoother and more coherent surfaces. 3DGAN appears to suffer from mode collapse on several categories, leading to lower coverage. Point clouds generated by PC-GAN are recognizable but lack detailed features; high-quality reconstruction from only 2048 generated points would be challenging. In addition, as shown in Figure~\ref{fig:GANinterp}, IM-GAN exhibits superior capability in 3D shape interpolation. As usual for the latent generative models, the interpolation is carried out by a linear interpolation between two latent codes; in-between 3D shapes are then generated from the intermediate codes.

We trained the IM-GAN further with $128^3$ resolution data on category plane, car and rifle. Figure~\ref{fig:teaser} shows some results sampled at $512^3$. We also include, in the supplementary material, videos showing interpolation results of IM-GAN sampled at $256^3$, and comparisons between interpolations in IM-AE and IM-GAN latent spaces.

\subsection{2D shape generation and interpolation}

\begin{table*}[!t]
\begin{center}
\begin{tabular}{l|c|c c|c c|c c|c}
\hline
 & DCGAN~\cite{DCGAN} & CNN-GAN & IM-GAN & VAE~\cite{VAE} & VAE$_{IM}$ & WGAN~\cite{WGANGP} & WGAN$_{IM}$ & Oracle \\
\hline\hline
COV-CD (\%) & 3.9 & {\bf 82.7} & 75.2 & 72.1 & {\bf 74.9} & \underline{{\bf 86.5}} & 84.7 & 88.4 \\
MMD-CD & 0.846 & 0.155 & {\bf 0.151} & 0.145 & \underline{{\bf 0.14}} & 0.158 & {\bf 0.149} & 0.137 \\
PWE (nat)& -282.83 & -8.07 & {\bf -6.16} & 17.39 & \underline{{\bf 30.6}} & -24.54 & {\bf -4.17} & 18.99 \\
PWE-nb (nat)& -230.47 & {\bf 130.93} & 128.38 & 304.57 & \underline{{\bf 318.07}} & {\bf 97.32} & 93.1 & 241.19 \\
IS~\cite{ISforMNIST} & 3.26 & 8.79 & {\bf 9.36} & 9.09 & \underline{{\bf 9.42}} & 8.9 & {\bf 9.22} & 9.8 \\
IS-nb & 3.26 & 8.8 & \underline{{\bf 9.39}} & 7.58 & {\bf 8.28} & 8.95 & {\bf 9.22} & 9.88 \\
\hline
\end{tabular}
\end{center}
\caption{ {\em Quantitative evaluation for 2D shape generation.} Oracle is the results obtained by using the training set as the sample set. The metrics without suffix ``-nb'' are evaluated using binarized images. PWE and IS are the higher the better. The better results in each subgroup (latent-GAN, VAE, WGAN) are bolded, and the best results over all models are underlined.}
\label{table:2DGANresults}
\end{table*}

\begin{figure*}[!t]
\begin{center}
   \includegraphics[width=1.0\linewidth]{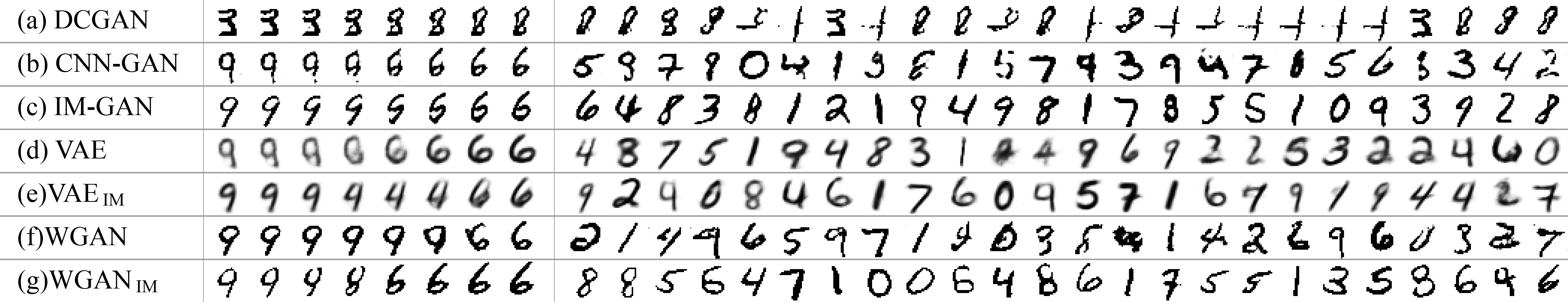}
\end{center}
   \caption{ {\em Visual results for 2D shape generation.} The first part of each row presents an interpolation between 9 and 6, except for DCGAN since it failed to generate number 6 or 9. The second part of each row shows some generated samples. The sampled images are not binarized. More samples and binarized images can be found in the supplementary material.}
\label{fig:2DGANvisual}
\end{figure*}

To evaluate IM-GAN for 2D shape generation, we conducted our experiments on the MNIST dataset since hand-written digits are naturally 2D shapes. We compared our results against DCGAN~\cite{DCGAN}, VAE~\cite{VAE}, and WGAN with gradient penalty~\cite{WGAN, WGANGP}. We also included the 2D version of CNN-GAN. In addition, we substituted the CNN decoders of VAE and WGAN with our implicit decoders, to obtain VAE$_{IM}$ and WGAN$_{IM}$. We trained all models on 5,000 binarized images from the training split of MNIST dataset for 1,000 epochs. The training set contains a smaller-than-usual amount of images, so that we can better observe the different features learned by CNN models and implicit models. The autoencoders of IM-GAN and CNN-GAN were pre-trained for 200 epochs. We replace LFD with chamfer distance in 2D images to obtain COV-CD and MMD-CD for evaluation. We also report the inception score for MNIST (IS)~\cite{ISforMNIST} and the log-likelihood produced by Parzen-window estimate (PWE)~\cite{parzen2, parzen1, parzen3}. For COV-CD and MMD-CD, we sampled 5,000 images from the subject models and compared against 1,000 ground truth images from the testing split. For IS and PWE, we sampled 10,000 images and used the entire testing split.

Quantitative and qualitative evaluations are shown in Table~\ref{table:2DGANresults} and Figure~\ref{fig:2DGANvisual}, respectively. Models equipped with our implicit decoders generally perform better. Due to inadequate training samples, DCGAN suffers from mode collapse, suggesting that the WGAN loss is preferred to extract true features with smaller training sets. VAEs have better performance over GANs when the output images are binarized, since VAEs tend to produce blurry results. For interpolation, CNN based methods tend to make old parts disappear and then new parts appear. This phenomenon is especially apparent in CNN-GAN and VAE. The implicit model usually warps the shape, but can also carry the ``disappear and appear'' trick. In visual comparison, IM-GAN and WGAN$_{IM}$ output cleaner and more recognizable ``in-between'' digits. One can find missing or redundant parts in the samples produced by CNN based methods, which are vestiges of the ``disappear and appear'' phenomenon.

\subsection{Single-view 3D reconstruction (SVR)}

\begin{table}[!t]
\begin{center}
\begin{tabular}{l|c|c|c|c|c}
\hline
  & Plane & Car & Chair & Rifle & Table \\
\hline\hline
HSP & 6,307 & 2,009 & 4,255 & 6,360 & 3,765 \\
AtlasNet25 & 4,877 & 1,667 & {\bf 3,244} & 6,507 & {\bf 2,725} \\
AtlasNetO & 5,208 & 1,751 & 4,124 & 6,117 & 3,909 \\
IM-SVR & {\bf 4,743} & {\bf 1,658} & 3,321 & {\bf 5,067} & 2,918 \\
\hline
\end{tabular}
\end{center}
\caption{ {\em Quantitative evaluation for SVR using LFD.}  The means are taken over the first 100 shapes in the testing set for each category and rounded to integers. AtlasNet25 is AtlasNet with 25 patches ($28,900$ mesh vertices in total) and AtlasNetO ($7,446$ vertices) is with one sphere. IM-SVR and HSP were both reconstructed at $256^3$ resolution. }
\label{table:SVRresults}
\end{table}

\begin{figure}[!t]
\begin{center}
   \includegraphics[width=1.0\linewidth]{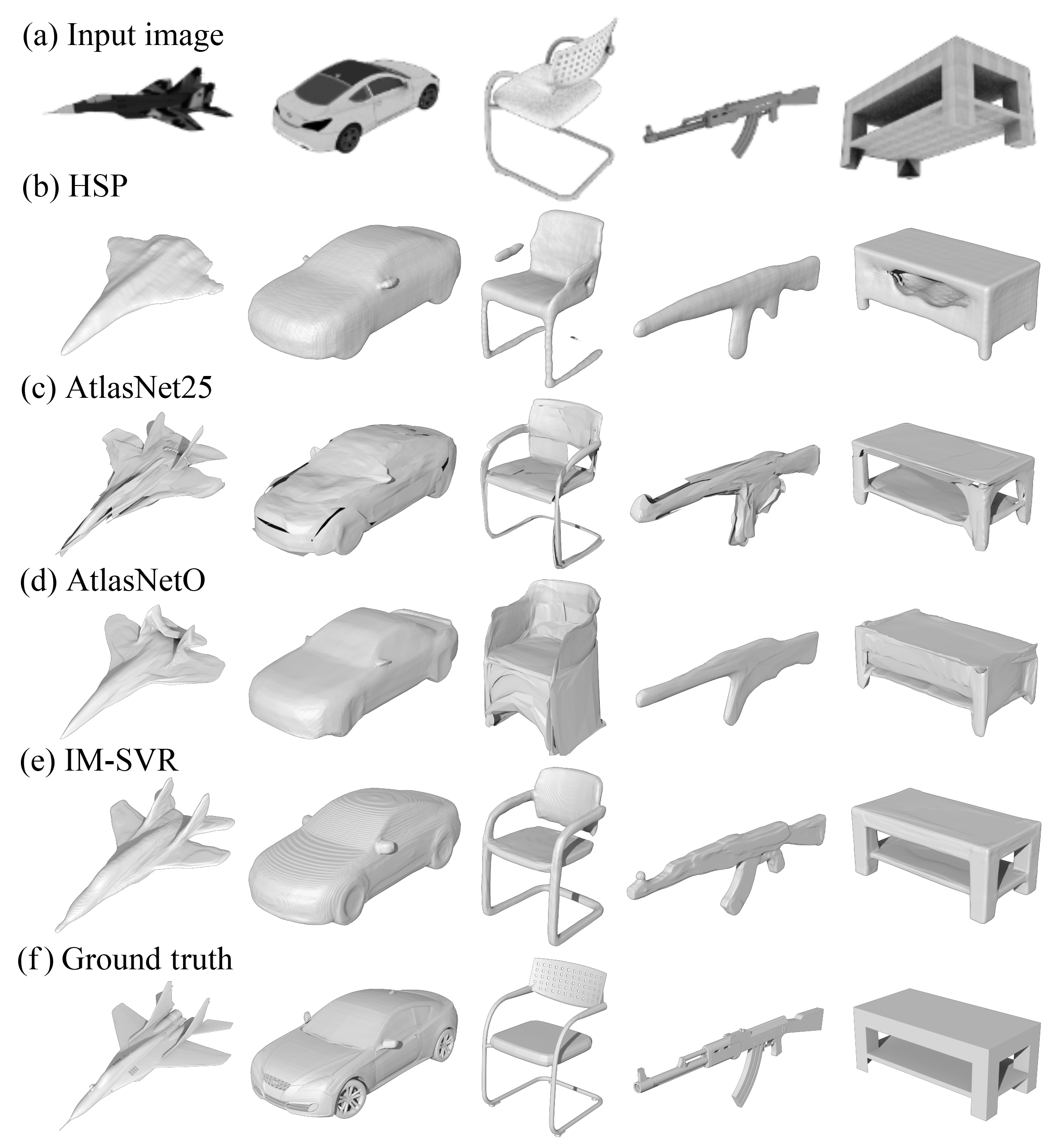}
\end{center}
   \caption{ {\em Visual results for single-view 3D reconstruction.} See caption of Table~\ref{table:SVRresults} for output model settings.} 
\label{fig:SVRvisual}
\end{figure}


We compare our approach with two state-of-the-art SVR methods, HSP~\cite{hsp}, an octree-based method that uses 3D CNN decoders to generate $256^3$ voxels, and AtlasNet~\cite{atlasnet}, which warps surface patches onto target shapes. For AtlasNet, we tested two setups for the initial surfaces, 25 patches or a sphere, and denote them by AtlasNet25 and AtlasNetO, respectively. For all methods, we trained individual models for each category, and used grayscale images as inputs. 

We used the train/test split in~\cite{hsp} to utilize the pre-trained models of HSP, since HSP requires quite a long time to converge. For HSP, we used the trained models made available online by the authors, and continued training for at most 2 days for each category and used the ones with the lowest testing loss. For AtlasNet, we trained the autoencoder part for 400 epochs and SVR part for 400 epochs and used the ones with the lowest testing loss. For our method, we trained IM-AE for 200-400 epochs on $64^3$ resolution, and IM-SVR for 1,000-2,000 epochs. The number of epochs depends on the size of the dataset. We did not train AtlasNet with such number of epochs since its testing loss had stopped dropping. Since IM-SVR was trained to map an image into a latent code, we did not have a good evaluation metric for testing errors. Therefore, we tested the last five saved checkpoints and report the best results.

Quantitative and qualitative evaluations are shown in Table~\ref{table:SVRresults} and Figure~\ref{fig:SVRvisual}, respectively. IM-SVR outputs were sampled at $256^3$ resolution, same as HSP. Output mesh settings for AtlasNet are from the authors' code. Though IM-SVR seems to have similar quantitative results with AtlasNet25, please keep in mind that LFD only captures the silhouette of the shape. AtlasNet25 represents shapes well yet clear artifacts can be observed since the shapes are made of patches and there is no measure to prevent slits, foldovers or overlapped surfaces. AtlasNetO can generate cleaner shapes than AtlasNet25, but the topology is pre-assigned to be equivalent to a sphere, thus AtlasNetO can hardly reconstruct shapes with holes. HSP can produce smooth surfaces but failed to recover most details. We show the comparisons of the first 16 shapes in the testing set for each category in the supplementary material, and also provide the evaluation results by chamfer distance to further verify that CD may not be an ideal evaluation metric for visual quality.

\section{Conclusion, limitation, and future work}

We introduced a simple and generic implicit field decoder to learn shape boundaries. The new decoder IM-NET can be easily plugged into contemporary deep neural networks for a variety of applications including shape auto-encoding, generation, interpolation, and single-view reconstruction. Extensive experiments demonstrate that IM-NET leads to cleaner closed meshes with superior visual quality and better handling of shape topology during interpolation.

A key merit of our implicit encoder is the inclusion of point coordinates as part of the input feature, but this comes as the cost of longer training time, since the decoder needs to be applied on each point in the training set. In practice, CNN-AE is typically 30 times faster than IM-AE on $64^3$ data without progressive training. Even with progressive training, IM-AE training took about a day or two and CNN-AE is still 15 times faster. When retrieving generated shapes, CNN only needs one shot to obtain the voxel model, while our method needs to pass every point in the voxel grid to the network to obtain its value, therefore the time required to generate a sample depends on the sampling resolution. While AtlasNet also employed MLP as decoder, AtlasNet25 is 5 times faster than ours in training, since AtlasNet only needs to generate points on the surface of a shape yet ours need to generate points in the whole field.

Our implicit decoder does lead to cleaner surface boundaries, allowing both part movement and topology changes during interpolation. However, we do not yet know how to regulate such topological evolutions to ensure a meaningful morph between highly dissimilar shapes, e.g., those from different categories. We reiterate that currently, our network is only trained per shape category; we leave multi-category generalization for future work. At last, while our method is able to generate shapes with greater visual quality than existing alternatives, it does appear to introduce more low-frequency errors (e.g., global thinning/thickening). 




In future work, we also plan to generalize IM-NET. First, using MLPs to decode may be too simple and inefficient; it is possible to improve the decoder structure to make the model size smaller for faster inference. Second, besides inside/outside signs, it is possible to train our decoder to output other attributes, e.g., color, texture, surface normal, signed distance, or deformation fields, for new applications. Finally, IM-NET has shown signs of understanding shape parts, suggesting its potential utility for learning part segmentation and correspondence.

\vspace{5pt}
\noindent
{\bf Acknowledgment.}~We thank Matt Fisher, Daniel Cohen-Or, and the anonymous reviewers for their comments, and Kangxue Yin and Ali Mahdavi-Amiri for proofreading. The research is supported by NSERC and an Adobe gift fund.



{\small
\bibliographystyle{ieee}
\bibliography{egbib}
}

\end{document}